\documentclass[aps,showkeys,showpacs,twocolumn,pre]{revtex4}
\usepackage{graphicx}
\usepackage{amsfonts}
\usepackage{graphicx}
\usepackage{graphics}
\usepackage{subfig}
\usepackage{bm}
\usepackage{dcolumn}

\def\bea{\begin{eqnarray}}
\def\eea{\end{eqnarray}}
\def\bphi{\bm{\phi}}
\def\bsigma{\bm{\sigma}}
\def\deepstrut{\vrule height 1.5ex depth 2.5ex width 0pt}
\def\rmd{\text{d}}
\def\rme{\text{e}}

\begin{document}
\title{Baby Skyrmions on the two-sphere}
\date{\today}

\author{Itay Hen}
\email{itayhe@post.tau.ac.il}
\author{Marek Karliner}
\email{marek@proton.tau.ac.il}
\affiliation{Raymond and Beverly Sackler School of Physics and Astronomy 
Tel-Aviv University, Tel-Aviv 69978, Israel}

\begin{abstract}
We find the static multi-soliton solutions 
of the baby Skyrme model on the two-sphere for topological charges
$1 \leq B \leq 14$.
Numerical full-field results show
that the charge-one Skyrmion is spherical, the charge-two Skyrmion is  
toroidal, and Skyrmions with higher charge all have point
symmetries which are subgroups of O(3).
We find that a rational map ansatz yields
very good approximations to the full-field solutions.
We point out a strong connection between the discrete symmetries of our solutions
and those of corresponding solutions of the $3D$ Skyrme model. 
\end{abstract}

\maketitle

\section{Introduction}
The Skyrme model \cite{Skyrme1} is a non-linear theory of pions in (3+1) dimensions
with topological soliton solutions called Skyrmions. 
The existence of stable solutions in this model is a consequence of the nontrivial
topology of the mapping $\mathcal{M}$ of the physical space
into the field space  at a given time,
$\quad \mathcal{M}:  S^3 \to SU(2) \cong S^3,\quad$ where 
the physical space $\mathbb{R}^3$  is compactified to $S^3$
by requiring the spatial infinity to be equivalent in
each direction. The topology which stems from this one-point 
compactification allows the classification of maps into equivalence classes,
each of which has a unique conserved quantity called the topological charge.
\par
Although Skyrmions were originally introduced to describe baryons in three 
spatial dimensions \cite{Skyrme1},
they have been shown to exist for a very wide 
class of geometries \cite{Geom}, and are
now playing an increasing role in other areas of physics as well.
\par
Apart from the original $3D$ model which exhibits very structured solitonic solutions (for a review see \cite{TopoSol}),
other Skyrme models are known to yield solutions with intricate structures. 
The baby Skyrme model, first introduced by \cite{Old1}, is
a two dimensional version of the original Skyrme model, with $\mathbb{R}^2$ as its domain. 
As its older brother, it is known to give rise,
under certain settings, to structured multi-Skyrmion configurations \cite{Old1,Old2,Holo1,Holo2,Holo3}.
Studies of other Skyrme models defined on curved domains, such as two-
and three-spheres can also 
be found in the literature \cite{2Sphere1,2Sphere2,3Sphere1,3Sphere2}.
Although most of these models are used as 
a simplification or as `toy' models of the full $3D$ model,
they also have physical significance on their own,
having several potential
applications in condensed-matter physics while often
revealing useful mathematical features.
\par
In the present paper we consider a baby Skyrme model on the two-sphere.
This type of model has been studied in \cite{2Sphere1,2Sphere2},
where only rotationally-symmetric configurations have been considered.
We compute the 
full-field minimal energy solutions of the
model up to charge $14$
and show that they exhibit complex multi-Skyrmion solutions
closely related to the Skyrmion solutions of the $3D$ model with the same
topological charge. 
To obtain the minimum energy configurations, we apply two completely different methods. 
One is a full-field relaxation method, with which exact numerical solutions of the model
are obtained. 
The other approach is a rational map approximation scheme,
which as we show yields very good approximate solutions. We discuss these methods
in detail in section \ref{sec:statSol}.
\par
In an exact analogy to the $3D$ Skyrme model,
our results show that the charge-one Skyrmion has a spherical energy distribution, 
the charge-two Skyrmion is  
toroidal, and Skyrmions with higher charge all have point
symmetries which are subgroups of O(3). 
The symmetries of these solutions are the same as those of
the $3D$ Skyrmions.
As we shall see, this is not a coincidence.

\section{The baby Skyrme model on the two-sphere}
The model in question is a baby Skyrme model 
in which both the domain and target are two-spheres.
It consists of a triplet of real scalar fields $\bphi=(\phi_1,\phi_2,\phi_3)$
subject to the constraint $\bphi \cdot \bphi =1$.
The Lagrangian density is simply
\bea \label{O3lagrangian}
\mathcal{L} &=& \frac1{2} \partial_{\mu} \bphi \cdot \partial^{\mu} \bphi \\\nonumber
&+& \frac{\kappa^2}{2} \big[ 
(\partial_{\mu} \bphi \cdot \partial^{\mu} \bphi)^2-
(\partial_{\mu} \bphi \cdot \partial_{\nu} \bphi)
(\partial^{\mu} \bphi \cdot \partial^{\nu} \bphi)
\big]\,,
\eea
with metric $\rmd s^2=d t^2 - \rmd \theta^2 -\sin^2 \theta \, \rmd \varphi^2$,
where $\theta$ is the polar angle $\in [0,\pi]$ and $\varphi$
is the azimuthal angle $\in [0,2 \pi)$.
The Lagrangian of this model
is invariant under rotations in both domain and the target spaces,
possessing an 
$\,O(3)_{\textrm{{\small domain}}} \times O(3)_{\textrm{{\small target}}}\,$ symmetry.
The first term in the Lagrangian is the kinetic term and
the second term, of the fourth order in derivatives, is the $2$D analogue of the Skyrme term \cite{Old1}. 
While in flat two dimensional space a third potential term is necessary to
ensure the existence of stable solutions, 
in the present  model it is not. 
Furthermore, stable 
solutions exist even with the Skyrme term missing. 
This is the well known $O(3)$ (or $\mathbb{CP}^1$) sigma model \cite{O3a}.\par
The field $\bphi$ in this model is an $S^2 \to S^2$ mapping. The relevant homotopy group 
of this model is $\pi_2(S^2)=\mathbb{Z}$, which implies that each field configuration is 
characterized by an integer topological charge $B$, the topological degree of
the map $\bphi$, which in spherical coordinates is given by
\begin{equation} \label{eq:O3Bdens}
B= \frac1{4 \pi } \int \rmd \, \Omega 
	\frac{ \bphi \cdot (\partial_{\theta} \bphi \times \partial_{\varphi} \bphi )}{
	\sin \theta} \,,
\end{equation}
where $\rmd \Omega= \sin \theta \, \rmd \theta \, \rmd \varphi$. 
\par
Static solutions within each topological sector are obtained by minimizing the energy 
functional
\bea \label{eq:O3energy}
E&=&\frac1{2} \int \rmd \Omega \Big((\partial_{\theta} \bphi)^2 
+ \frac{1}{\sin^2 \theta}( \partial_{\varphi} \bphi)^2 \Big) \\\nonumber
&+& \frac{\kappa^2}{2} \int \rmd \Omega \Big( \frac{(\partial_{\theta} \bphi \times \partial_{\varphi} \bphi)^2}{\sin^2 \theta}
\Big) \,.
\eea
Before proceeding, it is worth noting here that setting $\kappa=0$ 
in  Eq. (\ref{eq:O3energy}) yields the energy functional
of the $O(3)$ sigma model.
The latter has
analytic minimal energy solutions within every topological sector, given by
\begin{equation} \label{eq:rotSymForm}
\bphi=(\sin f(\theta) \cos(B \varphi),\sin f(\theta) \sin(B \varphi),\cos f(\theta))\,,
\end{equation}
where $f(\theta)=\cos^{-1}(1-2(1+(\lambda \tan \theta/2)^{2 B})^{-1})$ with $\lambda$ being some 
positive number \cite{O3a}.
These solutions are not unique, as other solutions with the same energy 
may be obtained by rotating (\ref{eq:rotSymForm}) either in the target or in the domain spaces. 
The energy distributions of these solutions in each sector are rotationally symmetric, with total
energy $E_{B}=4 \pi B$. \par
We have found that analytic solutions also exist for the energy 
functional (\ref{eq:O3energy}) with the Skyrme term only. 
They too have the rotationally symmetric form (\ref{eq:rotSymForm})
with $f(\theta)=\theta$ and total energy $E_{B}=4 \pi B^2$.
They can be shown to be the global minima
by the following Cauchy-Schwartz inequality:
\bea
&&\left( \frac1{4 \pi} \int \rmd \Omega \frac{ \bphi \cdot (\partial_{\theta} \bphi \times \partial_{\varphi} \bphi )}{
	\sin \theta} \right)^2 \\\nonumber
	&&\leq \left( 
	\frac1{4 \pi} \int \rmd \Omega \bphi^2
	\right) \cdot 	\left( \frac1{4 \pi} \int \rmd \Omega
	(\frac{\partial_{\theta} \bphi \times \partial_{\varphi} \bphi}{
	\sin \theta})^2
	\right) \,.
\eea
The LHS is simply $B^2$ and the first term in parenthesis 
on the RHS integrates to 1. 
Noting that the second term in the RHS is the Skyrme energy (without the $\kappa^2/2$ factor), 
the inequality
reads $E \geq 4 \pi  B^2$,
with equality for the rotationally-symmetric solutions.

\section{\label{sec:statSol} Static solutions}
In general, if both the kinetic and Skyrme terms are present,
static solutions of the model cannot be obtained analytically. 
The Euler-Lagrange equations
derived from the energy functional (\ref{eq:O3energy})
are non-linear \textit{PDE}'s, so the 
minimal energy configurations can only be obtained 
with the aid of numerical techniques. 
This is with the exception of the $B=1$ Skyrmion which has
an analytic ``hedgehog'' solution
\begin{equation}
\bphi_{[B=1]}=(\sin \theta \cos \varphi, \sin \theta \sin \varphi, \cos \theta)\,,
\end{equation}  
with total energy 
$\displaystyle \frac{E}{4 \pi}=1+\frac{\kappa^2}{2}\deepstrut$. 
\par
For Skyrmions with higher charge, we find the minimal energy configurations
by utilizing a full-field relaxation method, described in more detail below.
In parallel, we also apply the rational map approximation method, originally developed
for the $3D$ Skyrme model and directly compare the results with the relaxation method.
This method is also discussed below.

\subsection{Full-field relaxation method}
For the relaxation method, the domain $S^2$ is discretized 
to a spherical grid -- $100$ grid points for $\theta$ 
and $100$ points for $\varphi$. 
The relaxation process begins by initializing the field triplet $\bphi$
to a rotationally-symmetric configuration  
\begin{equation} \label{eq:initConfig}
\bphi_{\textrm{initial}}=(\sin \theta \cos B \varphi,\sin \theta \sin B \varphi,\cos \theta) ,
\end{equation}
where $B$ is the topological charge of the Skyrmion in question.
The energy of the baby Skyrmion is then minimized by repeating the following steps:
a point $(\theta_m,\varphi_n)$ on the grid is chosen at random, along 
with one of the three components of the field $\bphi(\theta_m,\varphi_n)$.
The chosen component is then shifted by a value $\delta_{\phi}$ chosen uniformly 
from the segment $[-\Delta_{\phi},\Delta_{\phi}]$
where $\Delta_{\phi}=0.1$ initially. The field triplet is then normalized
and the change in energy is calculated.
If the energy decreases, the modification of the field is accepted
and otherwise it is discarded.
The procedure is repeated while the value of 
$\Delta_{\phi}$ is gradually decreased throughout
the procedure. This is done until no further decrease in energy is observed.\par
One undesired feature of this minimization scheme
is that it can get stuck at a local minimum.
This problem can be resolved by using the ``simulated annealing'' algorithm \cite{SA1,SA2},
which in fact has been successfully implemented before,
in obtaining the minimal energy configurations of 
static two and three dimensional Skyrmions \cite{SkyrmeSA}. 
The algorithm comprises of repeated applications of a Metropolis algorithm 
with a gradually decreasing temperature,
based on the fact that when a physical system is slowly cooled down,
reaching thermal equilibrium at each temperature, it will end up in its ground state. 
This algorithm, however, is much more expensive in terms of computer time. 
We therefore employ it only in part,
just as a check on our results, which correspond to a Metropolis algorithm of zero temperature.
\par
As a further verification, we set up the minimization scheme using
different initial configurations and grids of different sizes ($80 \times 80$ and $200 \times 200$)
for several $\kappa$ and $B$ values.
This was done to make sure that the final configurations are independent of the
discretization and cooling scheme. 
Accuracy  was also verified by checking conservation
of the topological charge $B$ throughout the minimization process, yielding
$\displaystyle{\Big| \frac{B_{\textrm{{\scriptsize observed}}}}{B} - 1 \Big|< 10^{-6}}$.

\subsection{The rational map ansatz}
\par
Computing the minimum energy configurations 
using  the full non-linear energy functional is 
a procedure which is both time-consuming and resource-hungry.
To circumvent these problems, the rational map ansatz scheme has been devised.
First introduced in \cite{Rmaps3DSk1}, this scheme
has been used in obtaining approximate solutions
to the $3D$ Skyrme model using rational maps between Riemann spheres. 
Although this representation is not exact, it drastically 
reduces the number of degrees of freedom in the
problem, allowing computations to
take place in a relatively short amount of time. 
The results in the case of $3D$ Skyrme model are known
to be quite accurate.
\par
Application of the approximation, begins with expressing
points on the base sphere by the Riemann sphere coordinate
$\displaystyle z=\tan \frac{\theta}{2} \rme^{i \varphi}$. 
The complex-valued  function $R(z)$ is a rational map of
degree $B$ between Riemann spheres
\begin{equation}
R(z)=\frac{p(z)}{q(z)}\,,
\end{equation}
where $p(z)$ and $q(z)$ are polynomials in $z$, such that
$\max[\mbox{deg}(p),\mbox{deg}(q)]=B$,  and $p$ and $q$ have no common factors.  
Given such a rational map, the ansatz for the field triplet is
\begin{equation} \label{eq:Rmap}
\bphi=( \frac{R+ \bar{R}}{1+|R|^2}, i \frac{R- \bar{R}}{1+|R|^2}, \frac{1-|R|^2}{1+|R|^2})\,.
\end{equation}
It can be shown that rational maps of degree $B$ correspond to field configurations 
with charge $B$  \cite{Rmaps3DSk1}. Substitution of the ansatz (\ref{eq:Rmap}) into the 
energy functional (\ref{eq:O3energy}) results in the simple expression
\begin{equation} \label{eq:Ermap}
\frac1{4 \pi} E=B  + \frac{\kappa^2}{2} \mathcal{I} \,,
\end{equation}
with
\begin{equation} \label{eq:Iint}
\mathcal{I}=\frac{1}{4\pi}\int \bigg(
\frac{1+\vert z\vert^2}{1+\vert R\vert^2}
\bigg \vert\frac{d R}{d z}\bigg\vert\bigg)^4 \frac{2i \ \rmd z  \rmd \bar z
}{(1+\vert z\vert^2)^2}\,.
\end{equation}
Minimizing the energy (\ref{eq:Ermap}) only
requires finding the rational map which minimizes the functional
$\mathcal{I}$. As we discuss in the next section,
the expression for $\mathcal{I}$ given in Eq. (\ref{eq:Iint})
is encountered in the application of the rational map in
the context of $3D$ Skyrmions, where the procedure of 
minimizing $\mathcal{I}$ over all rational maps of the various degrees
has been used \cite{Rmaps3DSk1,Rmaps3DSk2,Rmaps3DSk3}. 
\par
Here we redo the calculations, using a relaxation method.
To obtain the rational map of degree $B$ that minimizes 
$\mathcal{I}$, 
we start off with a rational map of degree $B$, with the real and imaginary parts 
of the coefficients of $p(z)$ and $q(z)$ assigned random values from  the segment $[-1,1]$. 
As in the full-field relaxation method discussed above, solutions are obtained 
by relaxing the map until a minimum of $\mathcal{I}$ is reached.

\section{Relation to the 3D Skyrme model}
In the $3D$ Skyrme model,
the rational map ansatz can be thought of
as taken in two steps. 
First, the radial coordinate is 
separated from the angular coordinates by
taking the SU(2) Skyrme field $U(r,\theta,\varphi)$ to be of the form
\begin{equation} \label{ansatz}
U(r,\theta, \varphi)=\exp(if(r) \ \bphi(\theta,\varphi)\cdot \bsigma) \,,
\end{equation}
where $\mbox{\boldmath $\sigma$}=(\sigma_1,\sigma_2,\sigma_3)$ are Pauli matrices,
$f(r)$ is the radial profile function subject to
the boundary conditions  $f(0)=\pi$ and $f(\infty)=0$,
and $\bphi(\theta,\varphi): S^2 \mapsto S^2$ is a normalized vector 
which carries the angular dependence of the field.
In terms of the ansatz (\ref{ansatz}),
the energy of the Skyrme field is
\begin{widetext}
\begin{equation} \label{eq:energy3}
E=\int 4 \pi {f'}^2  r^2 \rmd r 
+\int 2({ f ' }^2 + 1)\sin^2 f \rmd r 
\int \Big( (\partial_{\theta} \bphi)^2 
+ \frac{1}{\sin^2 \theta}( \partial_{\varphi} \bphi)^2 \Big) \rmd \Omega
 +\int \frac{\sin^4f}{r^2} \rmd r 
 \int \frac{(\partial_{\theta} \bphi \times \partial_{\varphi} \bphi)^2}{\sin^2 \theta} \rmd \Omega.
\end{equation}
\end{widetext}
Note that the energy functional (\ref{eq:energy3})
is actually the energy functional of our model (\ref{eq:O3energy}) once the radial coordinate
is integrated out. Thus, our $2D$ model can be thought of as 
a $3D$ Skyrme model with a `frozen' radial coordinate.
\par
The essence of the rational map approximation is 
the assumption that $\bphi(\theta,\varphi)$ 
takes the rational map form (\ref{eq:Rmap}),
which in turn leads to a simple expression for the energy
\begin{equation} \label{eq:energy4}
E=4\pi\int \bigg(
r^2 {f'}^2+2B({f'}^2+1)\sin^2 f+\mathcal{I}\frac{\sin^4 f}{r^2}\bigg) \ \rmd r \,, 
\end{equation}
where $\mathcal{I}$ is given in Eq. (\ref{eq:Iint}).
As in our case, minimizing the energy functional 
requires (as a first step, followed by finding the profile function $f(r)$) minimizing
$\mathcal{I}$ over all maps of degree $B$.
\par
Since the symmetries of the $3D$ Skyrmions are determined solely by the angular dependence 
of the Skyrme field, it should not be too surprising that the solutions of the model discussed here
share the symmetries of the corresponding solutions of the $3D$ Skyrme model.

\section{Results}
The two approaches discussed in section \ref{sec:statSol}
have been applied separately to obtain the static
solutions for charges $2 \leq B \leq 14$ (the charge-one solution has an analytic representation,
as discussed in section \ref{sec:statSol}). 
This was done for several $\kappa$ values within the range $ 0.01 \leq \kappa^2 \leq 0.2$,
although solutions with different $\kappa$'s are
qualitatively similar. 
\par
\begin{figure*}
\subfloat[$B=2$]{
\label{figure1a} 
\includegraphics[angle=0,scale=1,width=0.23\textwidth]{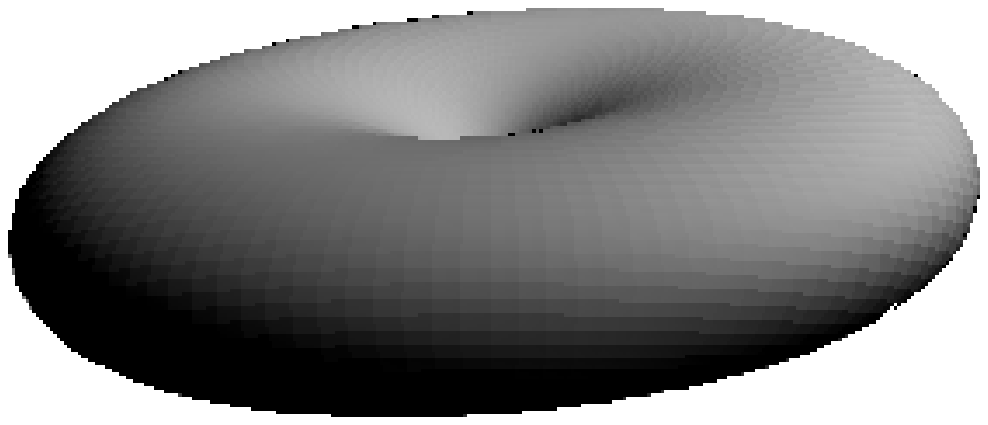}}
\subfloat[$B=3$]{
\label{figure1b} 
\includegraphics[angle=0,scale=1,width=0.23\textwidth]{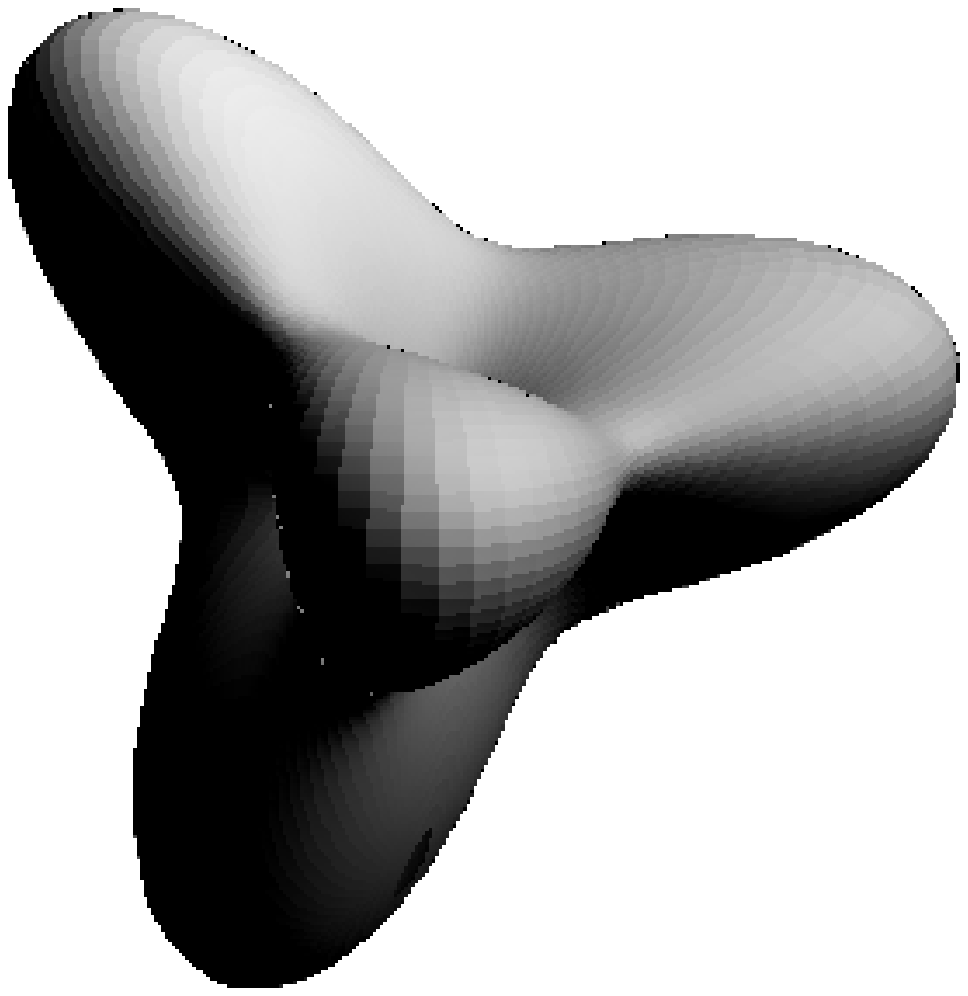}}
\hspace{1cm}
\subfloat[$B=4$]{
\label{figure1c} 
\includegraphics[angle=0,scale=1,width=0.23\textwidth]{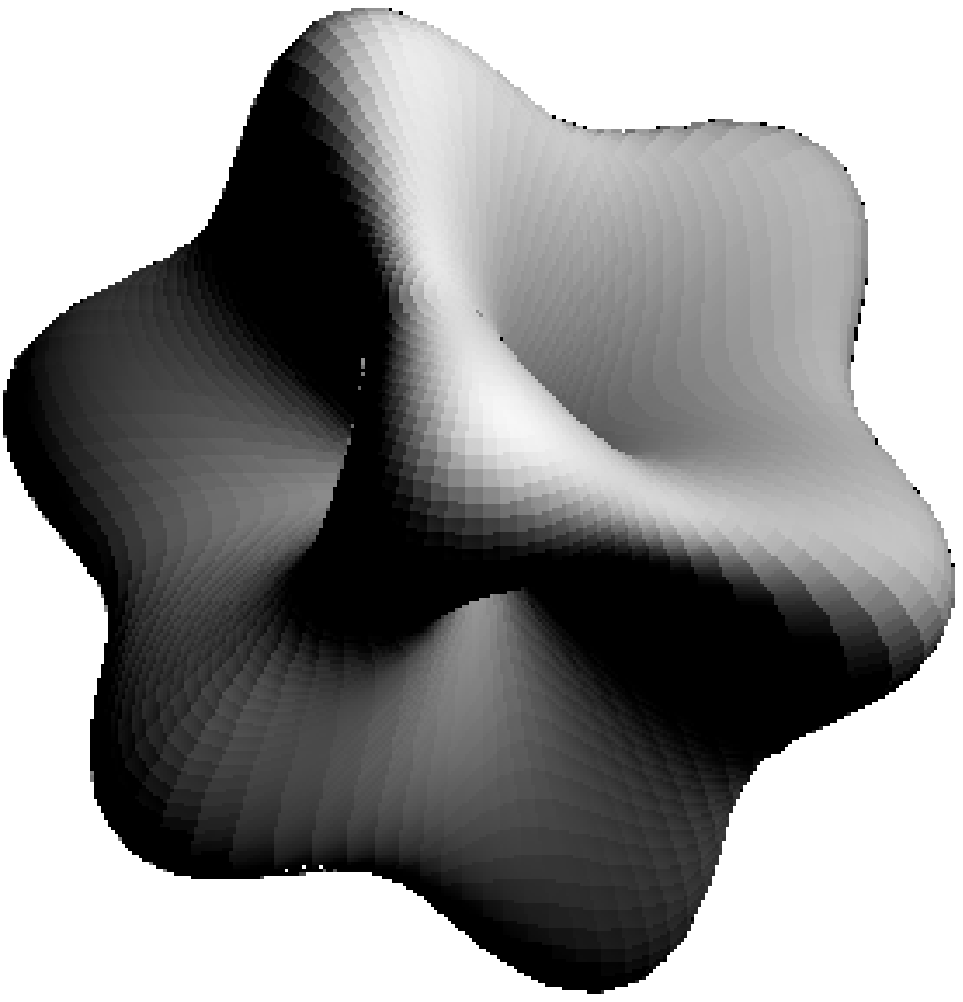}}
\subfloat[$B=5$]{
\label{figure1d} 
\includegraphics[angle=0,scale=1,width=0.23\textwidth]{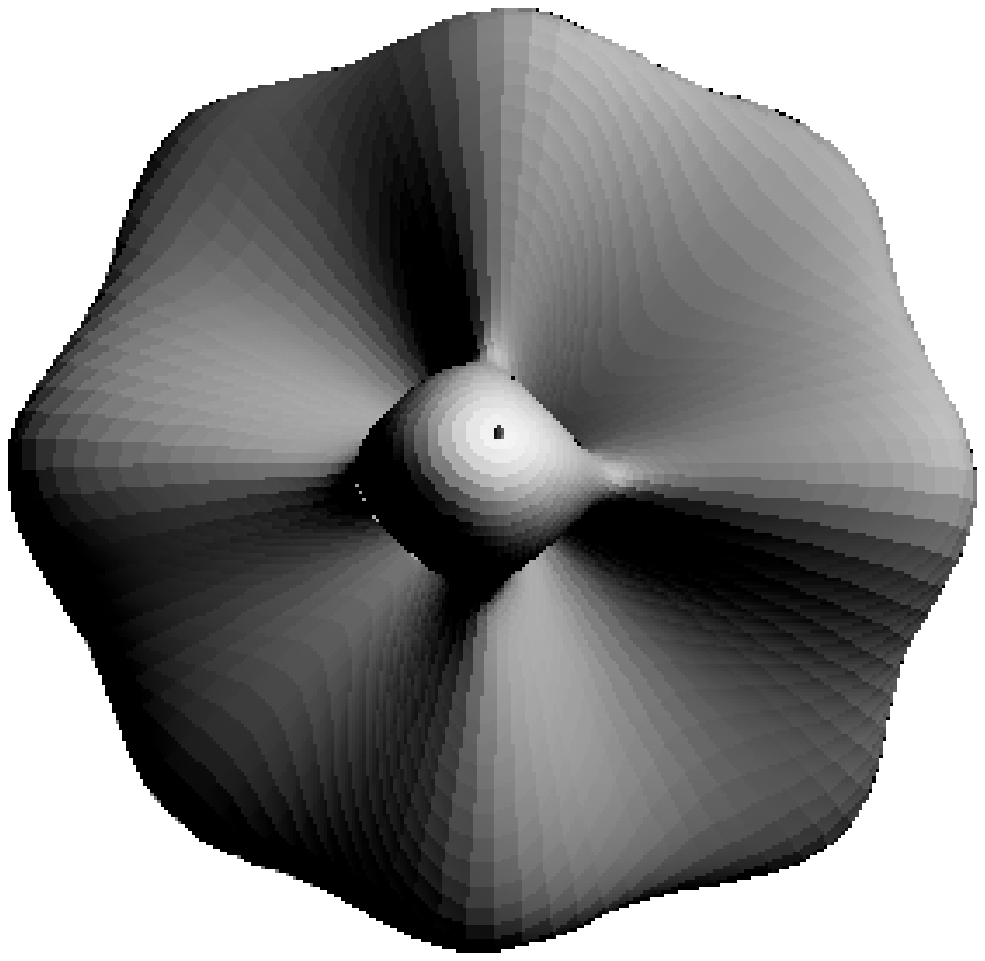}}
\subfloat[$B=6$]{
\label{figure1e} 
\includegraphics[angle=0,scale=1,width=0.23\textwidth]{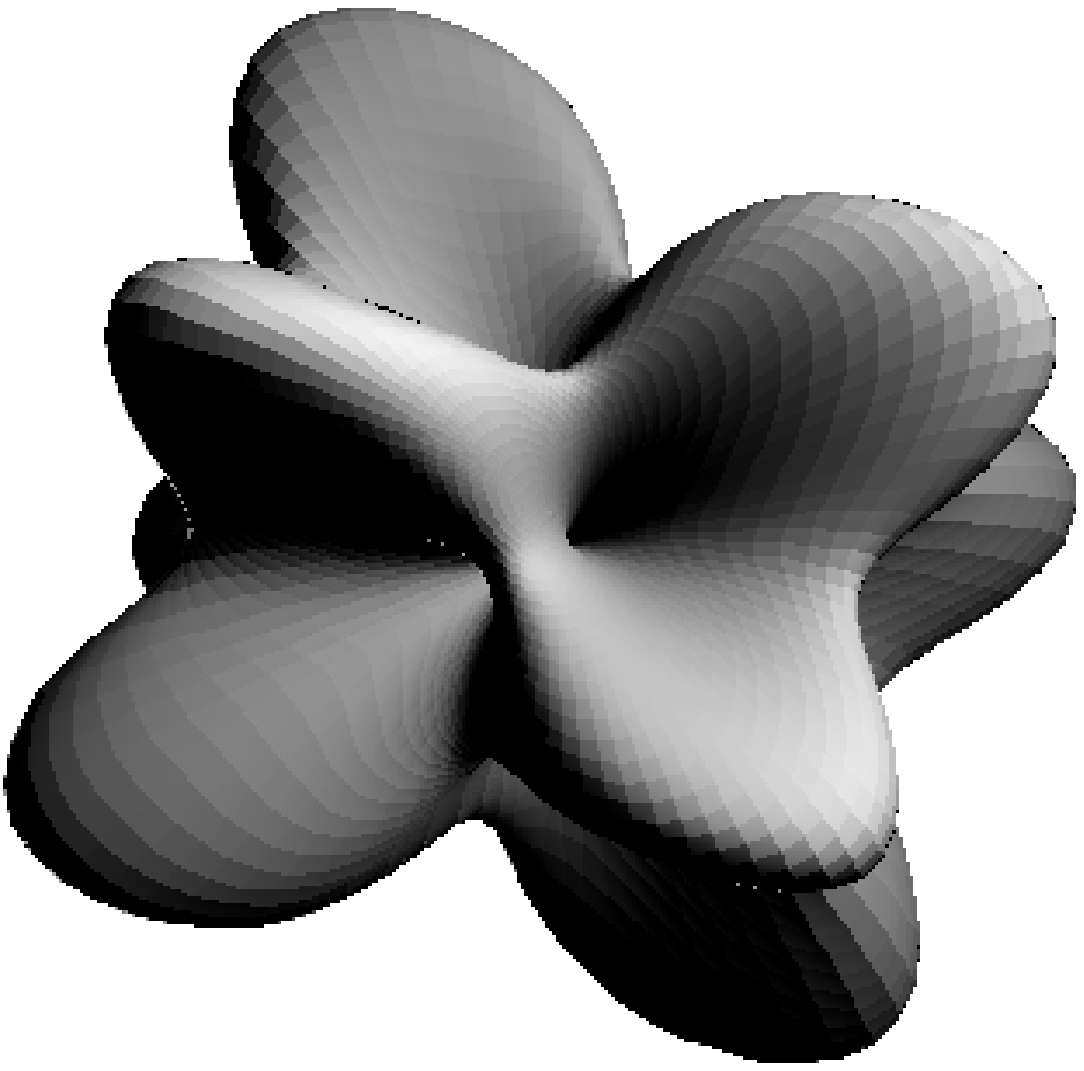}}
\hspace{1cm}
\subfloat[$B=7$]{
\label{figure1f} 
\includegraphics[angle=0,scale=1,width=0.23\textwidth]{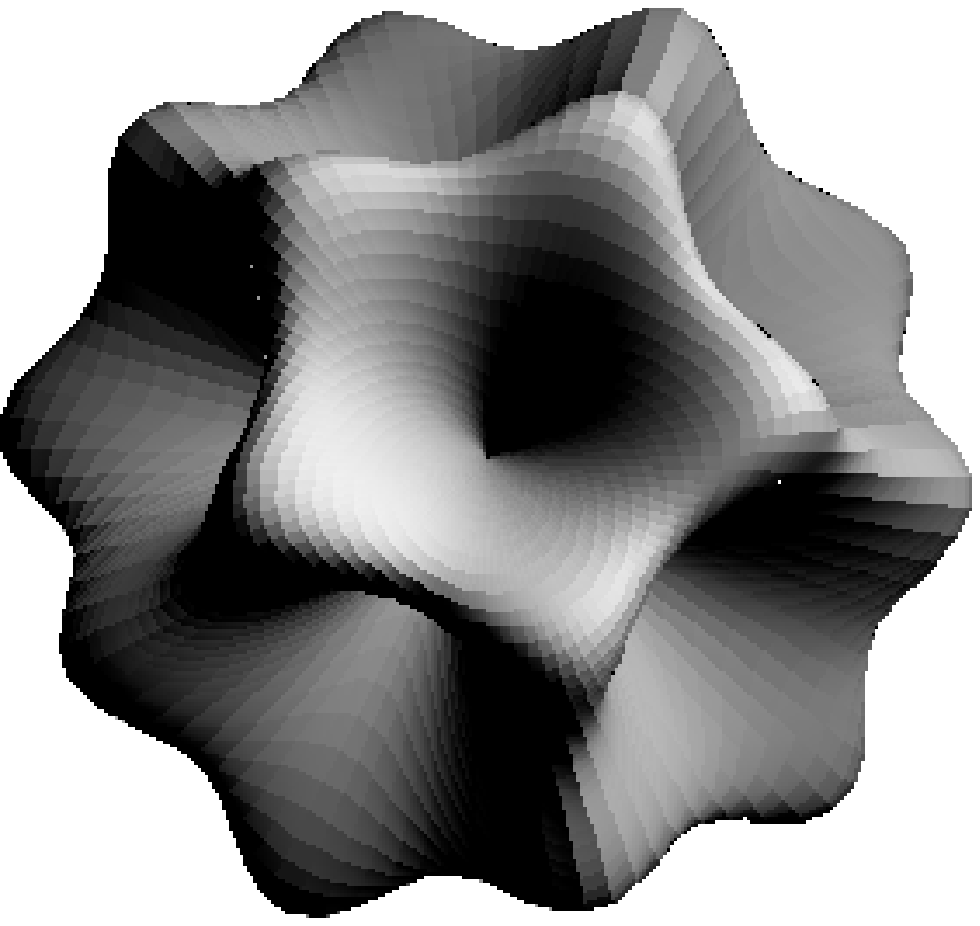}}
\subfloat[$B=8$]{
\label{figure1g} 
\includegraphics[angle=0,scale=1,width=0.23\textwidth]{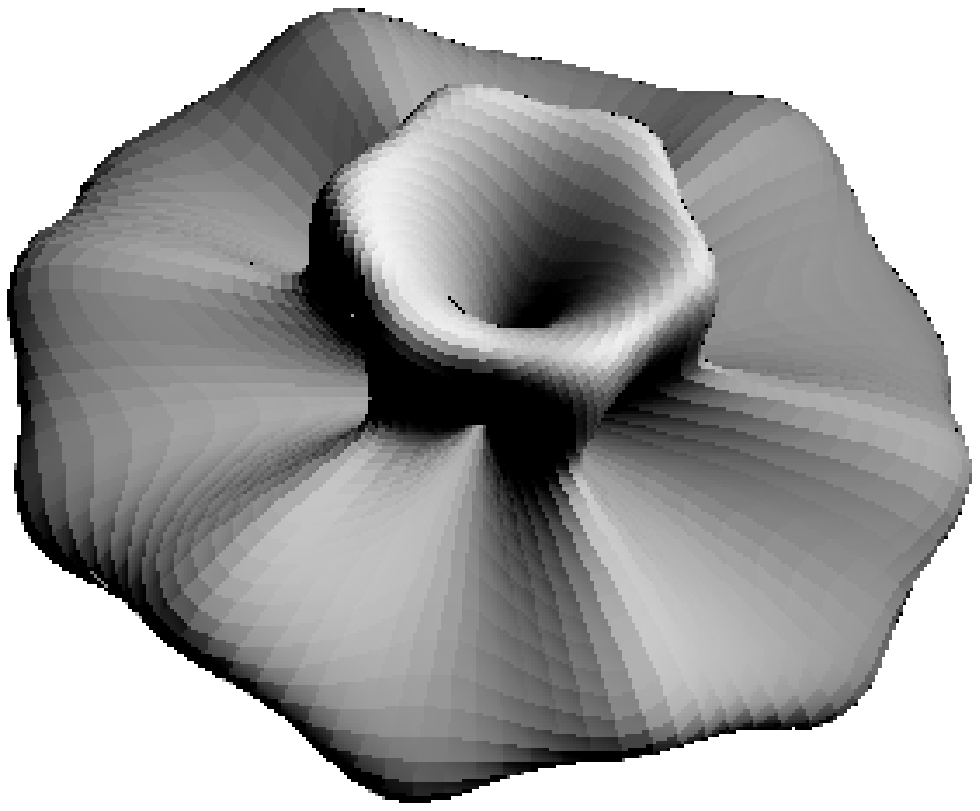}}
\subfloat[$B=9$]{
\label{figure1h} 
\includegraphics[angle=0,scale=1,width=0.23\textwidth]{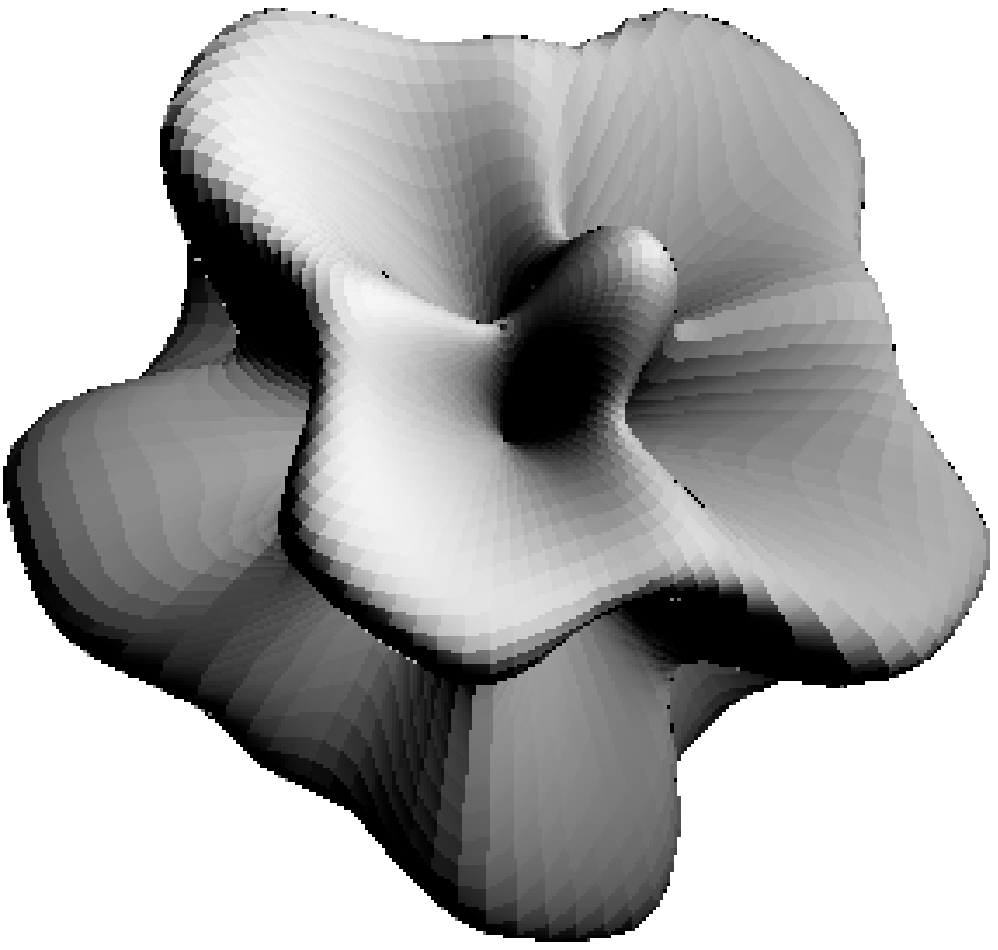}}
\hspace{1cm}
\subfloat[$B=10$]{
\label{figure1i} 
\includegraphics[angle=0,scale=1,width=0.23\textwidth]{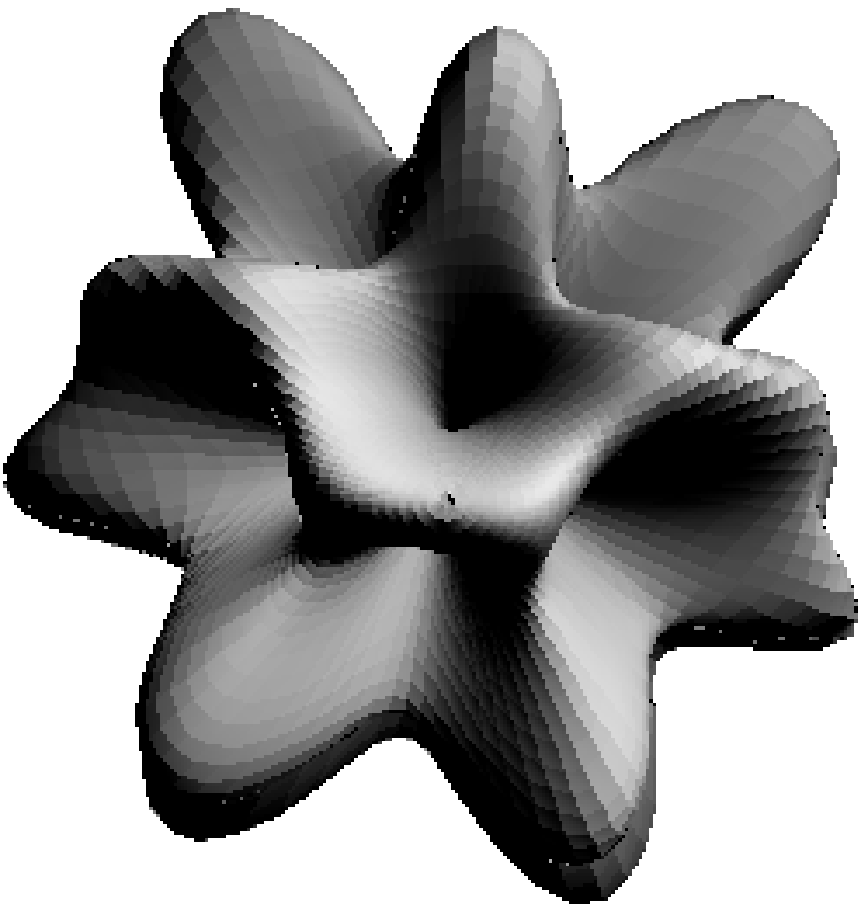}}
\subfloat[$B=11$]{
\label{figure1j} 
\includegraphics[angle=0,scale=1,width=0.23\textwidth]{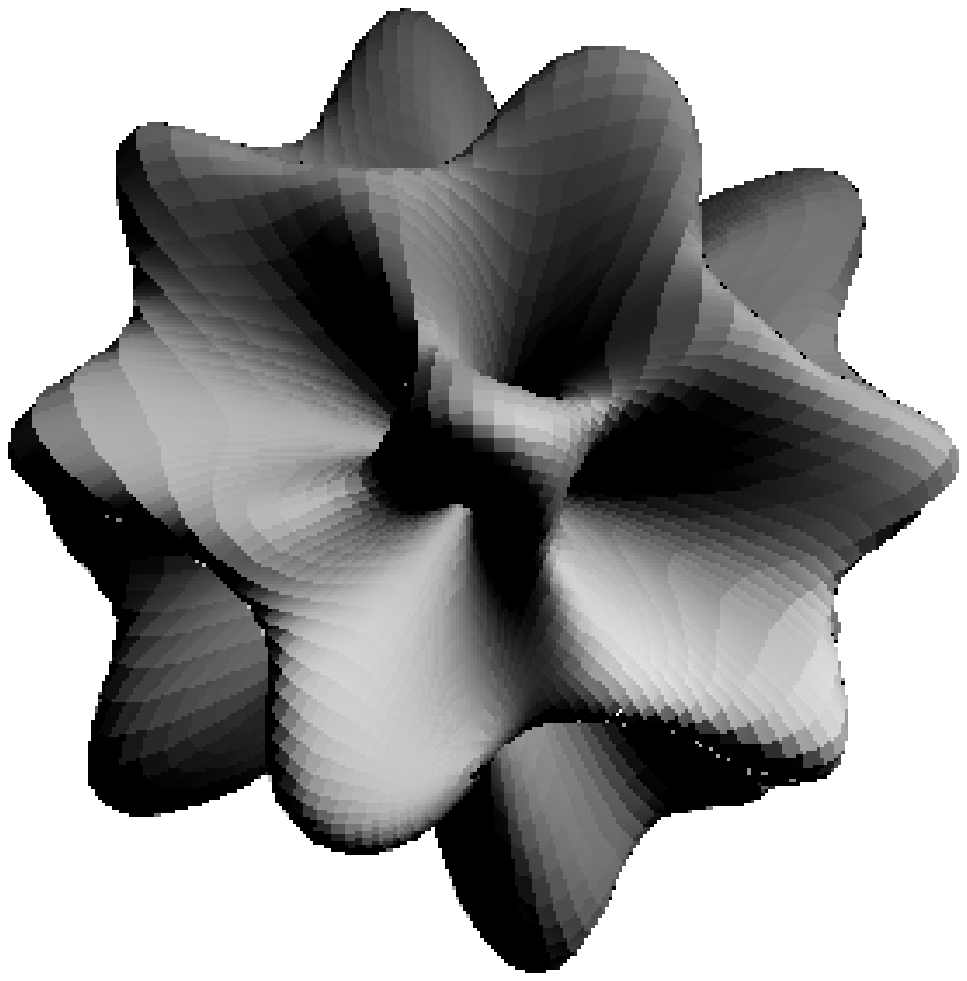}}
\subfloat[$B=12$]{
\label{figure1k}
\includegraphics[angle=0,scale=1,width=0.23\textwidth]{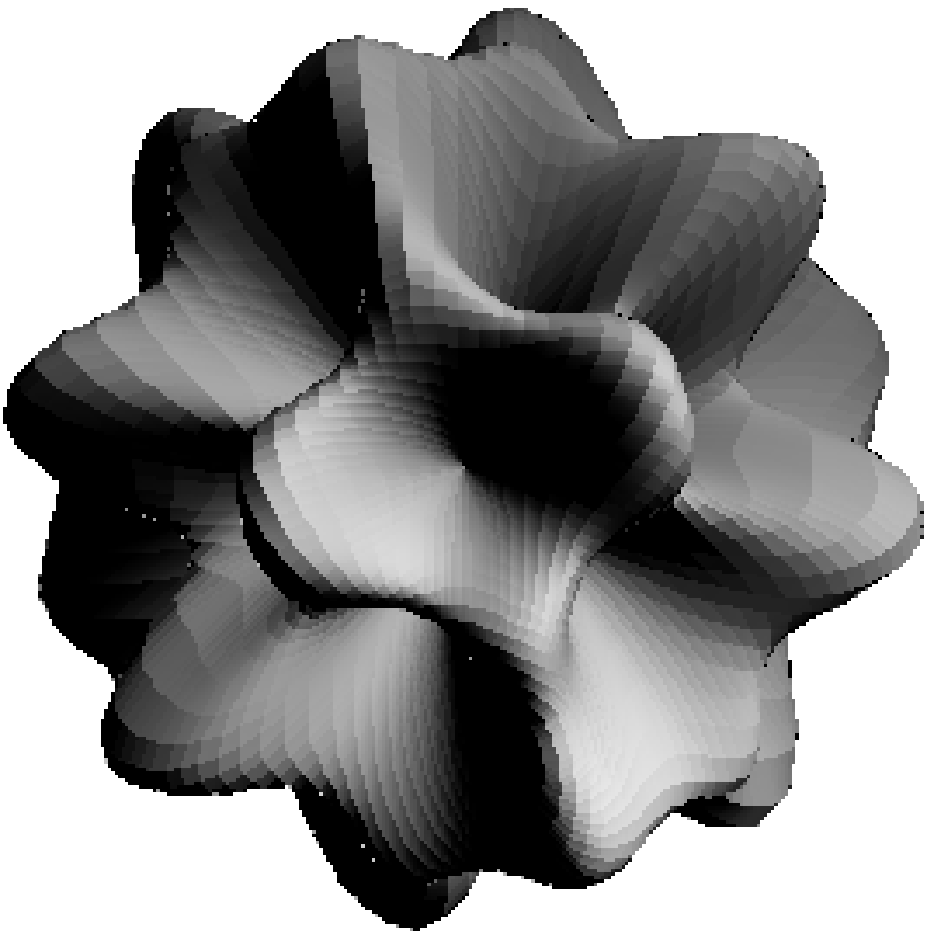}}
\hspace{1cm}
\subfloat[$B=13$]{
\label{figure1l} 
\includegraphics[angle=0,scale=1,width=0.23\textwidth]{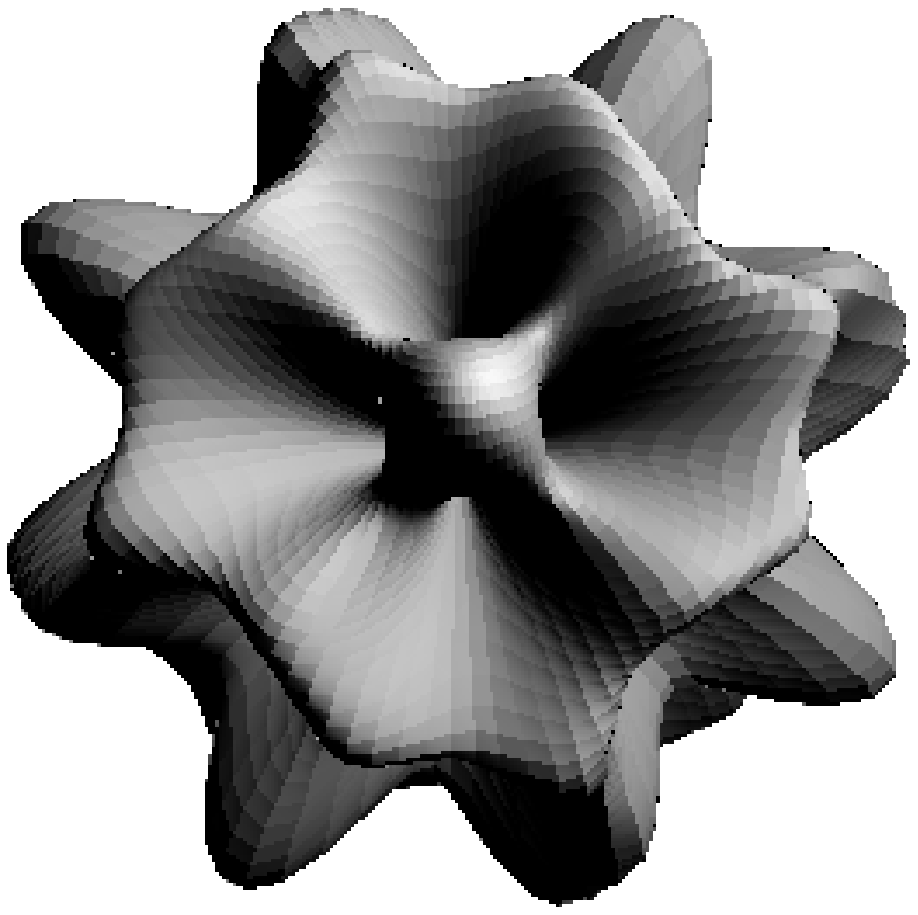}}
\subfloat[$B=14$]{
\label{figure1m} 
\includegraphics[angle=0,scale=1,width=0.23\textwidth]{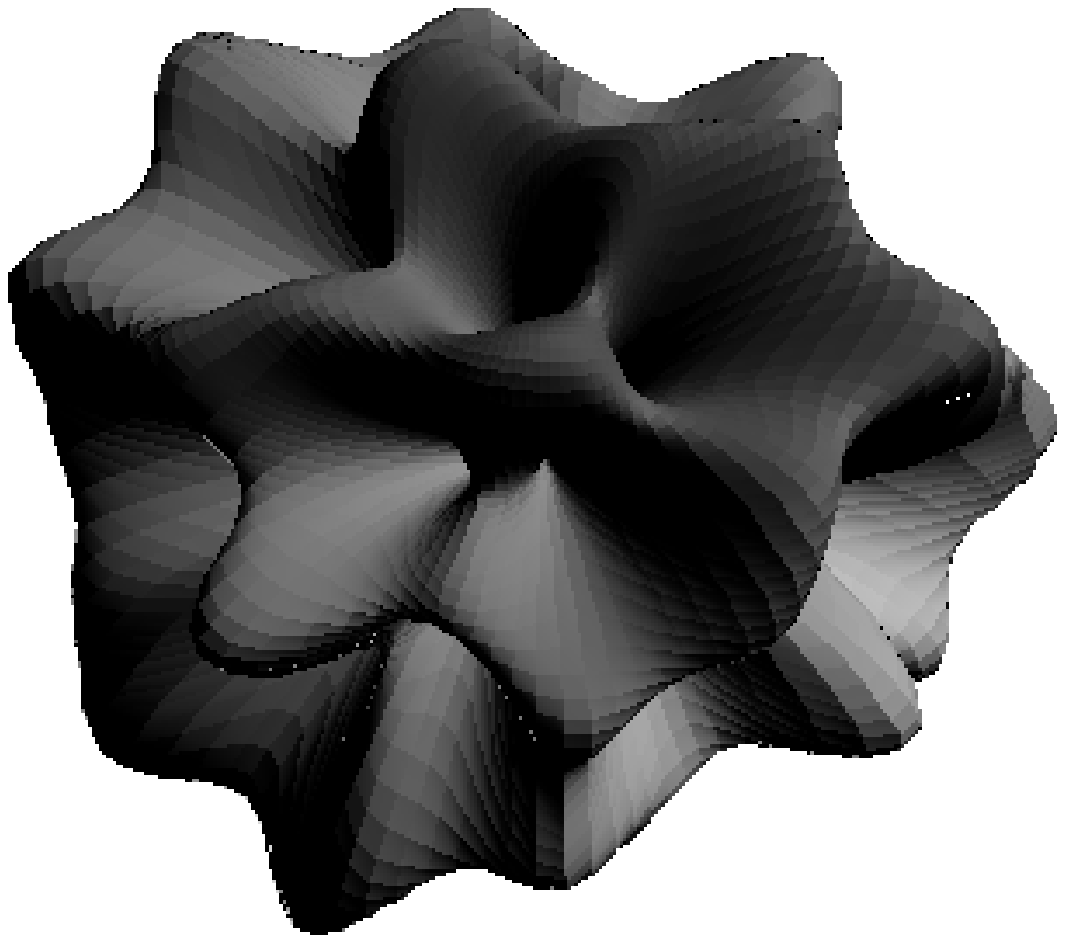}}
\caption{\label{Figure1}The energy distributions of the multi-Skyrmion solutions for charges \hbox{$2 \leq B \leq 14$} \hbox{($\kappa^2=0.05$).}}
\end{figure*}

As discussed in the previous section, the configurations obtained from the full-field relaxation method 
were found to have the same symmetries as 
corresponding multi-Skyrmions of the $3D$ model with the same charge.
The $B=2$ solution turned out to be axially
symmetric, whereas higher-charge solutions were all found to have point
symmetries which are subgroups of O(3). For $B=3$ and $B=12$,
the Skyrmions have a tetrahedral symmetry. 
The $B=4$ and $B=13$ Skyrmions have a cubic symmetry, 
and the $B=7$ is dodecahedral. 
The other Skyrmion solutions 
have dihedral symmetries. For $B=5$ and $B=14$ a $D_{2d}$ symmetry, 
for $B=6,9$ and $10$ a $D_{4d}$ symmetry, 
for $B=8$ a $D_{6d}$ symmetry and for $B=11$ a $D_{3h}$ symmetry.
The energy distributions of the obtained solutions for $\kappa^2=0.05$ are shown in 
FIG. ~\ref{Figure1}.
\par
While for solutions with $B<8$ the energy density (and also the charge
density) is distributed in distinct peaks,
for solutions with higher charge it is spread in a much more complicated manner.  
 
The total energies of the solutions (divided by $4 \pi B$) are listed in Table ~\ref{tab1}, 
along with the symmetries of the solutions (again with $\kappa^2=0.05$).
\par
Application of the rational map ansatz yielded results
with only slightly higher energies, only about $0.3 \%$ to $3 \%$
above the full-field results.
The calculated values of $\mathcal{I}$ were found to
agree with those obtained in \cite{Rmaps3DSk2} in the context of 
$3D$ Skyrmions. For $9 \leq B \leq 14$,
the rational map approximation yielded slightly less symmetric
solutions than the full-field ones.
Considering the relatively small number of degrees of freedom,
this method all-in-all yields very good approximations. 
The total energies
of the solutions obtained with the rational map approximation 
is also listed in Table ~\ref{tab1}.

\begin{table}[htp]
\caption{\label{tab1}Total energies (divided by $4 \pi B$)
of the multi-solitons of the model for $\kappa^2=0.05$.}
\begin{ruledtabular}
\begin{tabular}{@{}llllll}
Charge & Total energy & Total energy  & Difference & Symmetry of \\ 
$B$ & Full-field  & Rational maps & in $\%$    & the solution  \\ 
\hline
$2$  & $1.071$  & $1.073$  	& $0.177$      & Toroidal \\
$3$  & $1.105$  & $1.113$ 	& $0.750$      & Tetrahedral \\
$4$  & $1.125$  & $1.129$ 	& $0.359$ 	& Cubic \\
$5$  & $1.168$  & $1.179$ 	& $0.958$	& $D_{2d}$ \\
$6$  & $1.194$  & $1.211$ 	& $1.426$	& $D_{4d}$ \\
$7$  & $1.209$  & $1.217$ 	& $0.649$	& Icosahedral \\
$8$  & $1.250$  & $1.268$ 	& $1.406$	& $D_{6d}$ \\
$9$  & $1.281$  & $1.304$  	& $1.771$	& $D_{4d}$ \\
$10$ & $1.306$  & $1.332$ 	& $1.991$ 	& $D_{4d}$ \\
$11$ & $1.337$  & $1.366$ 	& $2.224$	& $D_{3h}$ \\
$12$ & $1.360$  & $1.388$ 	& $2.072$	& Tetrahedral \\
$13$ & $1.386$  & $1.415$ 	& $2.137$	& Cubic \\
$14$ & $1.421$  & $1.459$ 	& $2.712$	& $D_2$ \\
\end{tabular}
\end{ruledtabular}
\end{table}

\section{Summary and conclusion}
We have studied the baby Skyrme model on the two-sphere,
obtaining the minimal energy configurations for all charges up to
$B=14$, using both the full-field relaxation method 
and the rational map approximation scheme. 
For each charge we have identified the symmetry and
measured the energy of the minimal energy Skyrmion.
The solutions turned out to yield very structured configurations,
exhibiting the same symmetries as the 
corresponding solutions of the $3D$ Skyrme model. 
\par
We have explained the reason for these similarities between 
the symmetries of the two models and in the process we have exhibited
a strong connection between them.
The model discussed in this paper may be thought of as the $3D$ Skyrme 
model with a `frozen' radial coordinate.
In that sense, our computations may serve as an additional corroboration
of results obtained for the $3D$ Skyrmions. 
\par
In addition, we have shown that the rational map ansatz provides
a very good approximate description to the true solutions, also for high
topological charges.
The energies of the solutions computed in the rational maps approximation
are only slightly higher than the full-field solutions,
and their symmetries, in most cases, are the same.
This suggests that rational maps may be used 
to construct good approximations to multi-Skyrmion 
solutions in this model in a rather simple way.
\par
We believe that this work may provide a useful tool in the study of $3D$ Skyrmions,
as it shares great similarities with the $3D$ model, especially in terms of multi-Skyrmion symmetries.
The fact that the model discussed here is two-dimensional makes it 
simpler to study and perform computations with, when compared 
with the $3D$ Skyrme model. 

\begin{acknowledgments}
We thank Wojtek Zakrzewski for useful discussions. 
This work was supported in part by a grant from the Israel Science
Foundation administered by the Israel Academy of Sciences and Humanities.
\end{acknowledgments}

\end{document}